# Highly Deformable and Mobile Palladium Nanocrystals as Efficient Carbon Scavengers

Peng-Han Lu[1†], De-Gang Xie[1], Bo-Yu Liu[1], Fei Ai[1], Zhao-Rui Zhang[2], Ming-Shang Jin[2], Xiao Feng Zhang[3], Evan Ma[1,4], Ju Li[1,2,5*] and Zhi-Wei Shan[1*]

[1] Center for Advancing Materials Performance from the Nanoscale (CAMP-Nano) and XJTU-HHT Research & Development Center (XHRDC), State Key Laboratory for Mechanical Behavior of Materials, Xi'an Jiaotong University, Xi'an 710049, China

[2] Frontier Institute of Science and Technology, Xi'an Jiaotong University, Xi'an, Shaanxi 710054, China

[3] Hitachi High Technologies America, Pleasanton, CA 94588, USA

[4] Department of Materials Science and Engineering, Johns Hopkins University, Baltimore, MD 21218, USA

[5] Department of Nuclear Science and Engineering and Department of Materials Science and Engineering, Massachusetts Institute of Technology, Cambridge, Massachusetts 02139, USA

[†] Present address: Ernst Ruska-Centre for Microscopy and Spectroscopy with Electrons and Peter Grünberg Institute, Forschungszentrum Jülich, 52425 Jülich, Germany

*E-mail: liju@mit.edu; zwshan@mail.xjtu.edu.cn

Fouling of surfaces leads to performance degradation in many energy-intensive industrial processes, but the present solutions are either too complicated to be routinely used or incomplete for eradication. Here we propose and demonstrate that carbon-containing deposits can be catalytically wiped out in an efficient way by roaming palladium nanoparticles with extreme shape flexibility at relatively low temperatures. Surprisingly, during their dramatic liquid-like migrations, these particles could still maintain crystalline interior and conserve their initial crystal orientations through self-surface diffusion. Moreover, these catalytic particles were even able to become regenerated by other roaming particles after occasionally deactivated by surface coking or multiple-particle sintering. These findings shed light on metabolically driven, "living" nanocrystals, and also open a new avenue for efficient catalysis.

Carbon deposition on hot surfaces (e.g. in internal combustion engine, solid oxide fuel cell, electrochemical fuel production, etc.) usually leads to a measurable drop of energy efficiency and performance[1,2]. These carbon deposits originate from heat-induced decomposition of unconsumed vaporized fuel into solid fouling deposits. In a motor engine, components such as injectors, pistons, valves, spark plugs, throttle body, intake manifold, catalytic converter, and oxygen sensors etc., may all be potentially fouled by carbon[2]. Thus to bring the engine back to optimum combustion efficiency, routine maintenance is needed to clean up the fouling deposits, by adding chemical fuel additives[3,4] into the fuel or by immersing the components into ultrasonic baths. Adopting new engineering designs might also be helpful, but these are still incomplete solutions[5].

Catalytic oxidation could be an alternative way to eradicate fouling. However, distinct from many reactions where all reactant gases flow over anchored nanoparticles[6,7], catalytic defouling would involve a solid carbonaceous reactant besides gaseous oxygen. Given that carbon adheres to the surface and does not flow like gases, mobile catalytic nanoparticles would be necessary. Moreover, high shape flexibility is also essential for particles to stay adhered to the fouled surface while eating their ways forward. Finally, these particles must be able to maintain their catalytically active sites, such as certain crystalline facets, as they move.

The past decade has witnessed an explosion of elegantly synthesized metal nanocrystals with controlled sizes, shapes, compositions and structures[8,9]. Indeed, metal nanocatalysts often experience dynamic evolution in response to the surrounding gaseous environment, including sintering/redispersion[10,11], shape transition[12,13], surface reconstruction[14-17], surface segregation[18-20], phase transformation[21,22], etc. These dynamic changes usually correlate with their catalytic reactivity. For example, the morphology change of metal nanocrystals was found to bring about a transient[23] or oscillatory[24] reaction kinetics. Inspired by the latest findings that metal nanostructures can experience fast geometry transformation through surface diffusion[25,26], one may alternatively take advantage of these dynamic changes to mobilize catalytic nanoparticles and regenerate them after catalyst deactivation (e.g. coking and sintering)[27] so as to efficiently catalyze carbon removal.

Here we use palladium nanoparticles (Pd NPs) due to its demonstrated ability to catalyze the oxidation of graphite[28]. The solution containing Pd NPs of diameter ~10 nm were dispersed in ethanol under sonication and then dropped onto an amorphous $SiN_x$ support membrane. After drying up, a carbonaceous film, originating from the nonvolatile organic residues (e.g. surfactants, see more details in Supplementary Materials and Fig. S2), was left atop the surface of the support and some particles. This model system was then tested in a differentially pumped gas-environmental transmission electron microscope (Hitachi H-9500 ETEM) with a microchip-based heating stage, which can achieve direct visualization of the particle dynamics under oxygen atmosphere at elevated temperature. The schematic illustration of the experimental setup is shown in Supplementary Fig. S1.

The particles were firstly heated up to ~300 °C in vacuum. They stayed immobile even after half an hour except that the shape of some particles became a little round. Surprisingly, after a pure oxygen atmosphere was established around the sample with a partial pressure of ~0.2 Pa, some particles became "nanomotors" roaming around, but would become stationary again if the oxygen flow was cut off (see more details in Supplementary Materials Table S1). One typical example is illustrated in Fig. 1a-d. Fig. 1a, b displays the beginning (oxygen injection) and final (oxygen cut off) state of the particles, respectively. It can be seen that all of the six particles moved away from their initial positions. From the *in situ* movie data (Supplementary Movie S1), the moving trajectory of each of them was mapped out in Fig. 1c. These particles, however, did not show a universal moving direction, but seemed to move around semi-randomly like in Brownian motion. A check of the final state in the under-focus imaging mode demonstrated several etched channels on the substrate. Interestingly, those channels (Fig. 1d) coincided exactly with the trajectories of the particles extracted from the recorded movie (Fig. 1c), while in other areas where particles had not passed by, no such channels could be found. The channels were seen to be self-avoiding. That is, a particle would not cross the previous track of any other particles. Therefore, the random walks of these particles are not history-independent but has interesting non-Markovian (history-dependent) and multi-particle correlated features.

These dynamics of the nanoscale particle ensemble cannot be rationalized by equilibrium thermodynamics, but is driven by active chemical energy dissipation (energy metabolism). In order to unveil the mechanism of aforementioned phenomena, local chemical characterization using high angle annular dark field-scanning TEM (HAADF-STEM) imaging and electron energy loss spectroscopy (EELS) was performed around the channels. Fig. 1e presents a series of core-loss spectra collected across a channel shown in the inset image. The nitrogen (N) K peak deriving from the $SiN_x$ membrane existed in all the spectra, while the carbon (C) K peak could only be detected in spectra 1&4 (outside the channel). This demonstrated that the carbonaceous layer has been consumed during the particle migration. On the other hand, the nanobeam diffraction pattern of Pd NPs demonstrated that they remained face-centered cubic palladium structure after the reaction, instead of forming carbides or oxides (Supplementary Fig. S3). Therefore, it can be inferred that the carbonaceous layer over the SiNx substrate was oxidized into volatile carbon monoxide (CO) or carbon dioxide ($CO_2$) through the palladium-catalyzed reaction with oxygen at relatively low temperatures, and hence the carbon-deficient channels were etched out accompanying the migration of Pd NPs.

Apart from their flexible spatial mobility, we also found an elongation-contraction shape oscillation of these Pd NPs. Fig. 2a-d, are four typical snapshots from an *in situ* movie (Supplementary Movie S3) demonstrating the contraction (Figs. 2a and 2c) and elongation (Figs. 2b and 2d) in particle shape. In order to quantitatively describe the extent of deformation, the circumscribed ellipse of the particle was used to mimic its profile. The evolution of the eccentricity of the ellipse is plotted in Fig. 2e. It shows that the eccentricity quasi-periodically fluctuates between 0.65 (relatively round) and 0.93 (much more elongated), indicating a morphological oscillation of the particle. This thereupon gave rise to a peristalsis-fashion migration of the particle, which is akin to the movement of an earthworm by alternately shortening and lengthening of its body.

Furthermore, when imaged with electron beam out of focus, some shadows with bright contrast (red crosses in Fig. 3a-f) were seen displaced from the particle (red spots in Fig. 3a-f), and appeared intermittently at some fixed angles (Fig. 3g) during the particle migration (Supplementary Movie S6). The formation of these shadows resulted from the image delocalization of the corresponding particle at large defocus setting. The shifts of the delocalization relative to the particle are in the direction of the vectors of the strongly excited diffraction beams, and the distances in between are inversely proportional to the interspacing of the corresponding crystal plane[29]. It is striking to confirm that the polar coordinates of the shadows (red crosses in Fig. 3h) coincided well with the diffraction pattern of a face-centered cubic structure under [100] zone axis (blue dots in Fig. 3h). It presents a persuasive evidence to prove that the active particle, although behaving like liquids morphologically, always maintained a crystalline interior with unchanging/conserved crystal orientation (always travelling around the fixed zone axis with negligible off-axis angle during its migration). This suggests that the locomotion did not involve obvious rigid-body rotation, but was dominated by surface diffusion, which is also affirmed in a bi-grained particle with a pinned grain boundary (see Supplementary Fig. S5 and Movie S5). We have thus established a high-throughput technique to simultaneously monitor the morphology and crystallography of the nanocrystals during their locomotion by performing *in situ* TEM with under-focus imaging condition.

Physically, the particle movements are accomplished by Coble creep, that is, surface diffusion of Pd atoms, which happens so rapidly at this length scale that it is sufficient to sustain the mass transport needed for the shape changes. The key difference with traditional Coble creep is that it is driven by chemical energy release instead of mechanical energy release or capillary energy reduction[25]; that is, the Pd atoms diffuse to where C+O*→CO↑ or $CO_2$↑ can happen. Consequently, the Pd particle is then motivated to spread part of itself towards the neighboring carbon, protruding in a way that allows it to meet carbon to mediate further catalytic reaction. This results in the observed "elongation", as the ends appear as if they were anchored and "stretched" by the carbon borderline that is receding away from the particle. However, the catalytic etching perpendicular to the carbon borderline might experience different speeds at each point, especially at both ends of the particle due to the perturbation of the surface diffusion of Pd atoms parallel to the borderline. As a result, the slope of the carbon borderline and thus the elongation direction of the particle were modified constantly, as indicated by the angles between the horizontal line and one fixed end of the long axis of the ellipse which are plotted in Fig. 2f. On the other hand, the particles do not elongate indefinitely, and would sometime shrink towards a more spherical shape driven by surface energy minimization. This competition between active catalytic metabolism of the surroundings and self-energy minimization provokes the shape oscillation and "peristaltic" migration of Pd NPs. Especially, the shape oscillations of the particles enabled themselves to eat out carbon-free passages much longer and wider than their equilibrium sizes, which enhanced the efficiency of the catalytic carbon removal.

Two main effects that could lead to the deactivation of the catalysts needs to be discussed further. The first one is surface coking[30], which refers to the blocking of the metal surface by the accumulation of carbon. Since the catalytic oxidation is initiated by the chemisorption of oxygen on the Pd surface, a dense surface carbonaceous layer over the NPs would serve as a kinetic barrier to prevent the oxygen from directly contacting the catalyst, thus poisoning the catalytic reaction. However, these barricades are not all uniform or dense. Some particles with bald spots on the surface barrier layer could adsorb the oxygen, facilitate the metabolism of carbon over the particle surface (see the *in situ* high resolution ETEM results in Supplementary Fig. S4) as well as around the particle, and thus become the "pioneers" moving around (e.g. Fig. 1a-d). Moreover, a mobile NP will also, upon contact, help to remove the protective carbon shell of an initially-dormant particle that it touches, as illustrated in Fig. 4e-h. As a result of this "contagious mobilization", more particles will be activated upon encountering mobile particles, which will then start to move themselves (Fig. 4a-d, Supplementary Movie S2). This is thus a self-propelling chain reaction of mobilization and decoking.

The other effect is particle sintering. In fact, we found that particle coalescence can be counteracted statistically by particle splitting, analogous to the cell fusion and fission in biology. One typical example is shown in Fig. 5 (Supplementary Movie S4). Three particles coalesced and then underwent shape elongation. However, soon after, this coalesced particle split into two, like in biological cell division. This splitting is an alternative process to the particle contraction: there seems to be a critical neck thickness below which the polarizing influence of the carbon on the two ends is too great to maintain unity. The split particles were then able to move around as independent entities.

As either the density of the NPs increases, or the carbon resources nearby get exhausted, the NPs will migrate less, split off less and eventually turn into immobile particles that slowly coarsen, and the system will demonstrate less "living" but more near-equilibrium features. We conjecture based on our understanding of the dynamics, however, that if one were able to replenish carbon on the substrate continuously as in a fuel injection engines or many actual industrial processes, then fusion and fission events of NPs might balance statistically, and we should achieve a NP population at quasi-steady state, because the principle of maximum entropy production rate in non-equilibrium thermodynamics calls for many small NPs with long total carbon-Pd contact lines to metabolize carbon, rather than a gigantic fused particle favored by equilibrium thermodynamics.

In summary, we have discovered an efficient catalytic carbon removal system with coking- and sintering-resistant nanocatalysts that maintains crystalline interior with strongly conserved crystallographic orientation, working via peristalsis-like migration with high shape flexibility and spatial mobility through surface diffusion. This lifelike dynamics of inorganic matter represents a startling emergent behavior that can arise out of an energy-metabolizing nano-system, and may inspire more efficient use of nanostructures that are coarsening-resistant in energy-intensive industrial processes.

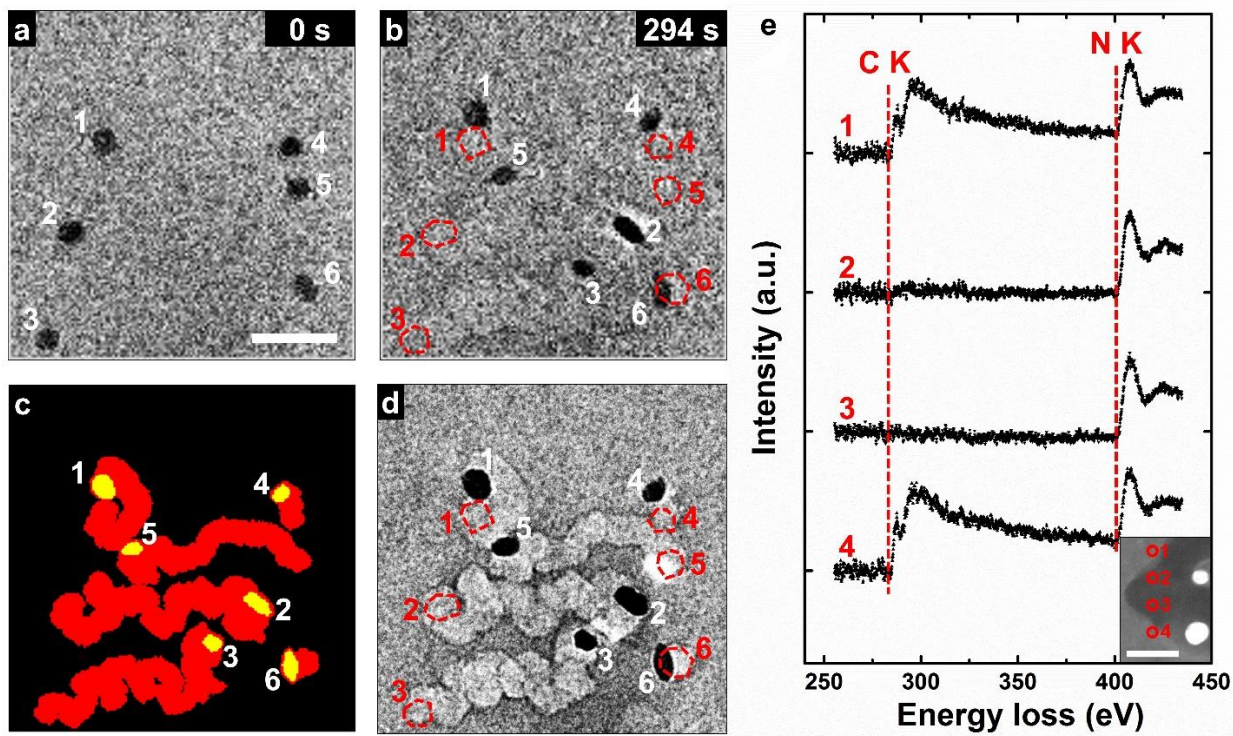

**Figure 1.** "Self-propelled" migration. a-d, "Self-propelled" migration of Pd NPs under 0.2 Pa pure oxygen at 300 oC. The moving trajectories (c) of particles from the initial (a, also red circles in b and d) to the final (b) state coincided with the channels that were found in the final state at under-focused imaging mode (d). Scale bar, 50 nm. See Supplementary Movie S1. e, The electron energy loss spectra collected across a carved channel showed that the carbon inside the channel was consumed after the particle passing through. The collecting positions of each spectrum were marked in the inset HAADF-STEM image. Scale bar, 50 nm.

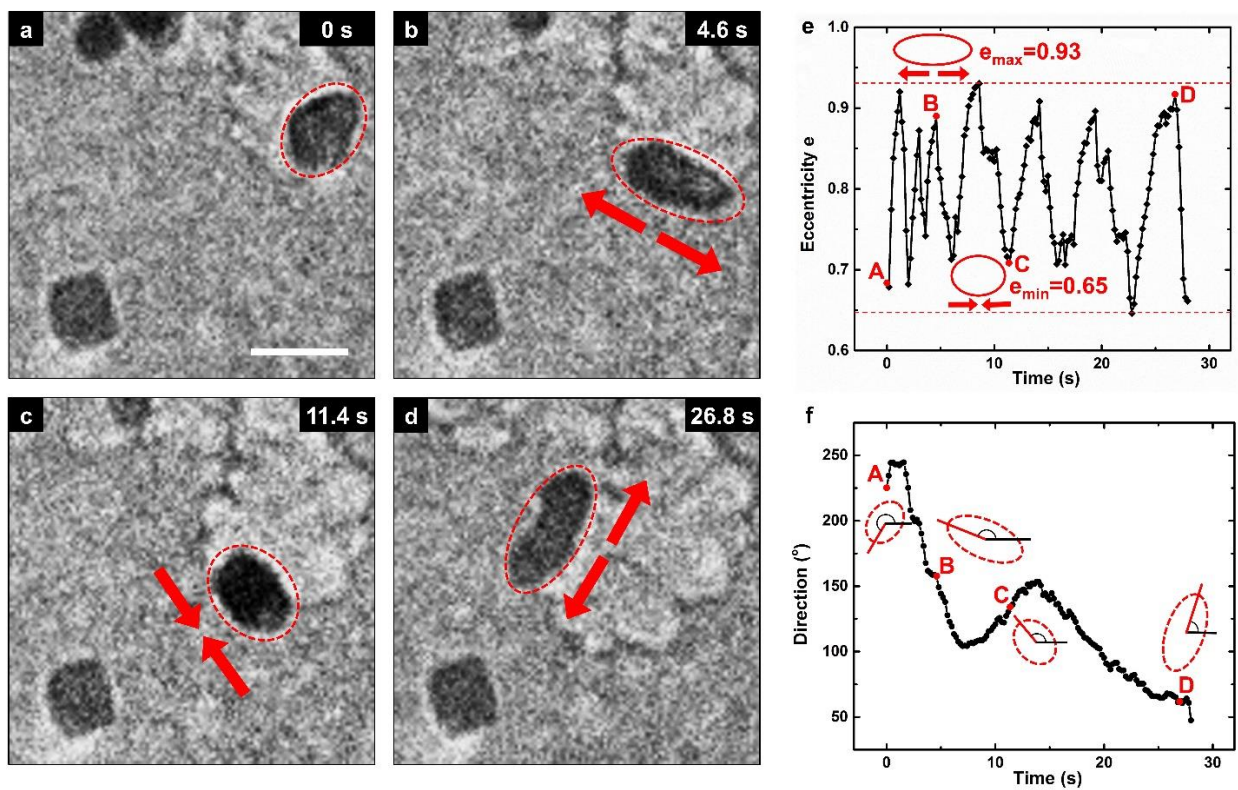

**Figure 2.** Shape oscillation and peristalsis-like locomotion. a-d, *In situ* TEM snapshots illustrating an elongation-contraction shape oscillation and thus peristalsis-like locomotion of a Pd NP. Scale bar, 20 nm. See Supplementary Movie 3. e, Time-dependent evolution of the eccentricity of the particle geometry as delineated by the circumscribed ellipses in (a-d). f, Time-dependent elongation direction of the particle. The direction is denoted by the angle between the horizontal line and one fixed end of the long axis of the ellipse. The four insets correspond to (a-d).

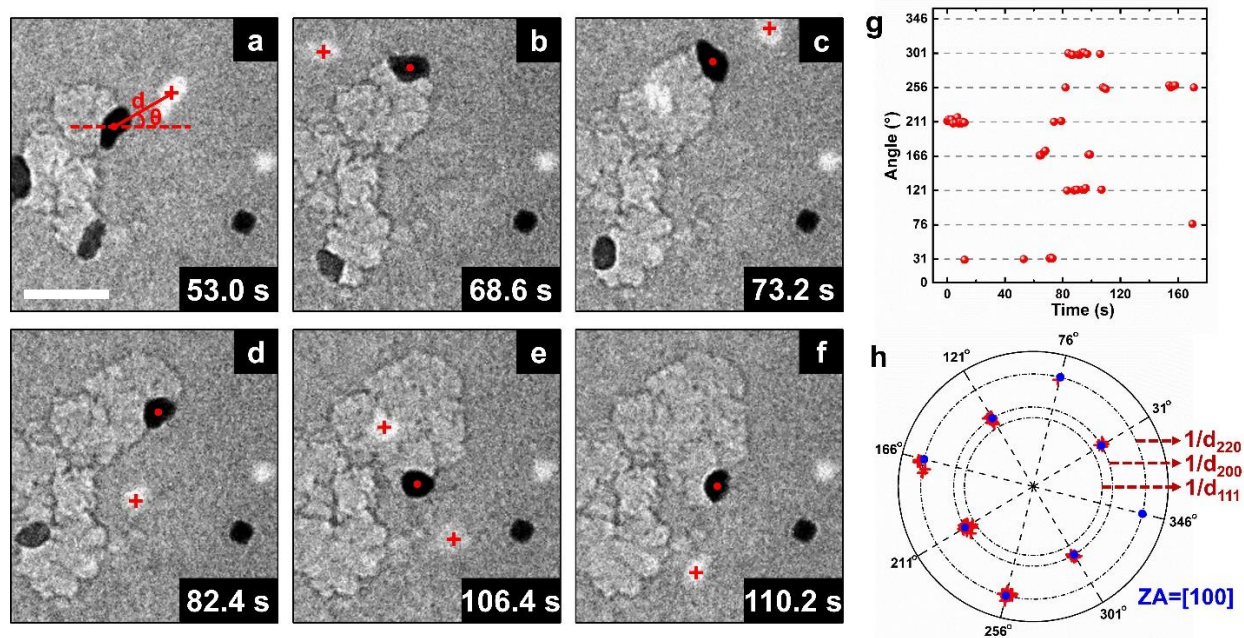

**Figure 3.** Crystalline orientation-conservation. a-f, Six example snapshots from an in situ movie in the under-focus imaging mode. The shadow images with bright contrast could be found near the particles. Scale bar, 50 nm. g, Time-dependent azimuthal angle of the shadow image marked with red "+" symbol in a-f.  h, Polar coordination diagram of the azimuth angle and relative distance between the shadow image and the particle from each snapshot (red crosses). They coincided very well with the diffraction pattern of a face-centered cubic structure under [100] zone axis (blue dots). See Supplementary Movie S6.

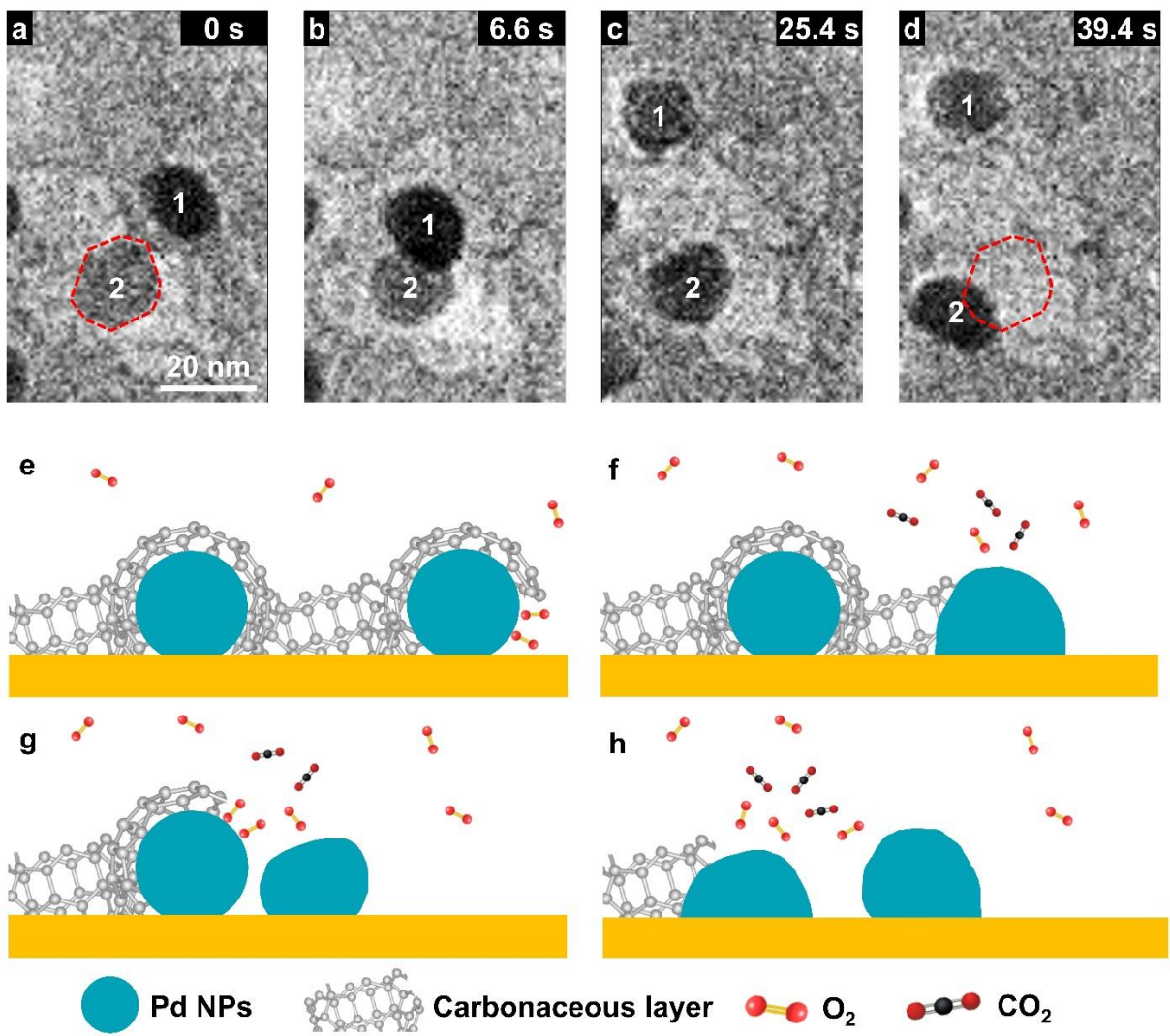

**Figure 4.** Surface decoking and contagious mobilization. a-b, Time-lapsed TEM images of Pd NPs showing a "contagious mobilization" behavior. After the contact with the mobile Particle 1, Particle 2 (delineated by dashed red polygon in a and d) started migrating away from its initial position (c, d). Scale bar, 20 nm. See Supplementary Movie S2. e-h, Schematic illustration of the chain-reaction resulted from the surface decoking by contiguous particles.

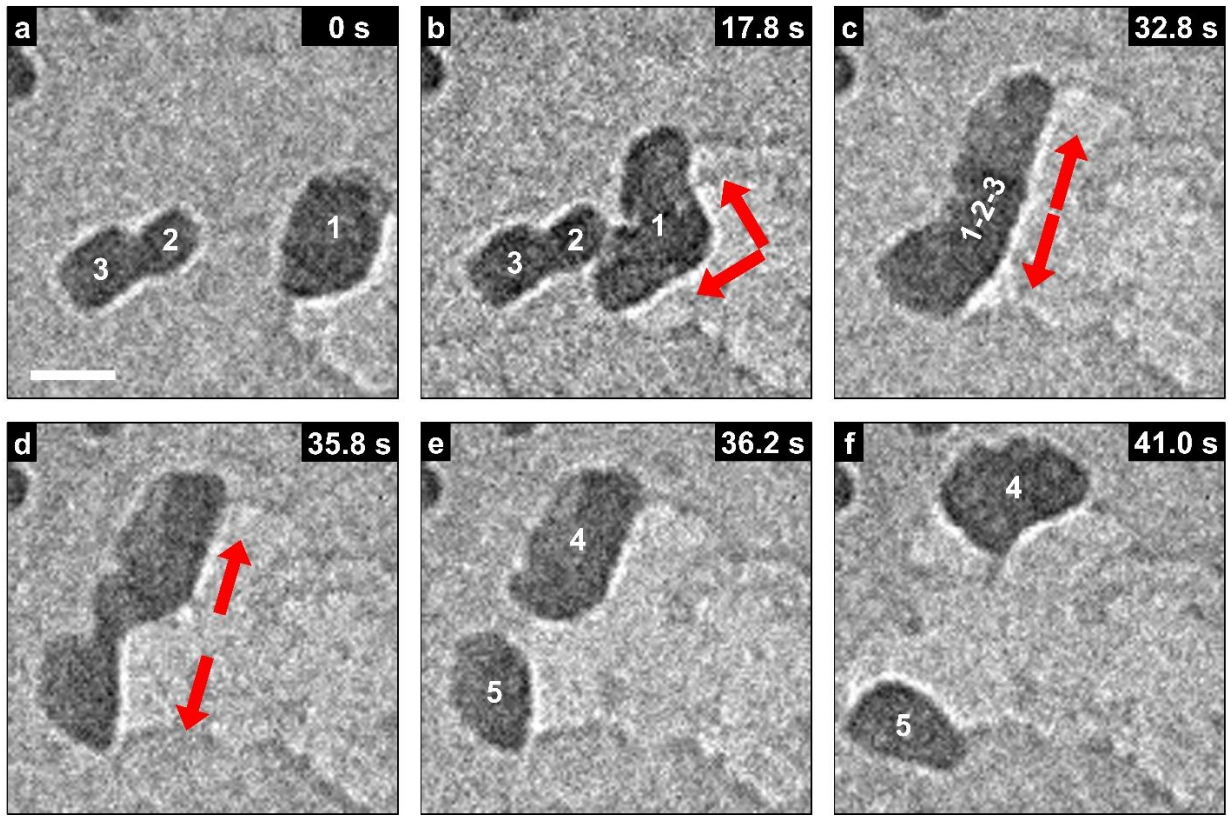

**Figure 5.** Particle fission after fusion. a-c, Fusion of the particles. Particle 1 moved around and coalesced with Particles 2&3 into a larger one. d-f, Fission of a larger particle and the split particles were then able to move around as independent entities. All scale bar, 20 nm.

**Supporting Information**

Supporting information of nanoparticle synthesis, TEM experimental setup, Table S1, Figure S1-S5 and *in situ* TEM movies S1-S6.

**Acknowledgment**

The authors acknowledge support from the Natural Science Foundation of China (51231005, 51401159 and 51321003), and 973 Programs of China (2012CB619402). We also appreciate the support from the 111 project (B06025). J.L. acknowledges support by NSF CBET-1240696 and DMR-1410636. E.M. acknowledges support from US DoE-BES-DMSE, under Contract No. DE-FG02-13ER46056. We also thank Dr. Chao Wu (Xi'an Jiaotong University, Xi'an, China) and Prof. Kaili Jiang (Tsinghua University, Beijing, China) for helpful discussions.

**Author Contributions**

Z.-W.S. and J.L. supervised the project. P.-H.L. performed the *in situ* ETEM experiments. B.-Y.L. and P.-H.L. conducted the STEM-EELS characterization. P.-H.L., D.-G.X. and F.A. analyzed the data with input from Z.-W.S., J.L., E.M. and X.F.Z. Z.-R.Z. and M.-S.J. synthesized the nanoparticle samples. P.-H.L., J.L., E.M. and Z.-W.S. wrote the manuscript. All the authors discussed the results and commented on the manuscript.

Supporting Information for

# Highly Deformable and Mobile Palladium Nanocrystals

# as Efficient Carbon Scavengers

Peng-Han Lu[1†], De-Gang Xie[1], Bo-Yu Liu[1], Fei Ai[1], Zhao-Rui Zhang[2], Ming-Shang Jin[2], Xiao Feng Zhang[3], Evan Ma[1,4], Ju Li[1,2,5*] and Zhi-Wei Shan[1*]

[1] Center for Advancing Materials Performance from the Nanoscale (CAMP-Nano) and XJTU-HHT Research & Development Center (XHRDC), State Key Laboratory for Mechanical Behavior of Materials, Xi'an Jiaotong University, Xi'an 710049, China

[2] Frontier Institute of Science and Technology, Xi'an Jiaotong University, Xi'an, Shaanxi 710054, China

[3] Hitachi High Technologies America, Pleasanton, CA 94588, USA

[4] Department of Materials Science and Engineering, Johns Hopkins University, Baltimore, MD 21218, USA

[5] Department of Nuclear Science and Engineering and Department of Materials Science and Engineering, Massachusetts Institute of Technology, Cambridge, Massachusetts 02139, USA

[†] Present address: Ernst Ruska-Centre for Microscopy and Spectroscopy with Electrons and Peter Grünberg Institute, Forschungszentrum Jülich, 52425 Jülich, Germany

*E-mail: liju@mit.edu; zwshan@mail.xjtu.edu.cn

**Synthesis of Pd nanoparticles.** Pd cubic seeds with an average edge length of 10 nm were synthesized following a previously reported protocol[1]. In a typical synthesis experiment, 8.0 mL of deionized (DI) water (18.2 M$\Omega$·cm), 105 mg of Poly(vinyl pyrrolidone) (PVP, MW$\approx$55,000, Aldrich), 60 mg of L-ascorbic acid (AA, Aldrich), and 300 mg of KBr (Aldrich) were mixed together in a 20 mL vial and pre-heated in air under magnetic stirring at 80 $^\circ$C for 10 min. 57 mg of $Na_2PdCl_4$ (Aldrich) were dissolved in 3 mL of DI water and then introduced into the pre-heated solution. After the vial had been capped, the reaction was allowed to proceed at 80 $^\circ$C for 3 h. The final product was collected by centrifugation.

***In situ* heating ETEM experiments.** The *in situ* heating experiments were carried out on a micro-electro-mechanical system (MEMS) thermal chip (Protochips Aduro heating holder). Pd NPs were dispersed in ethanol under sonication and dropped onto a MEMS chip for the *in situ* TEM experiments. They were directly heated up to 300 $^\circ$C at a heating rate of 2 $^\circ$C/s and kept at that temperature for ~30 min. Pure oxygen gas up to 0.2 Pa was then injected into the TEM specimen chamber. The dynamics of Pd NPs under oxygen atmosphere were investigated with Hitachi H-9500 environmental transmission electron microscope (ETEM) operated at 300 keV. The microscope is featured by a differentially pumped system and thus can accommodate reactive gases within the specimen chamber. The real-time TEM movies were recorded with Gatan Orius 832 high-speed charge coupled device (CCD) camera at a time resolution of 0.2 s.

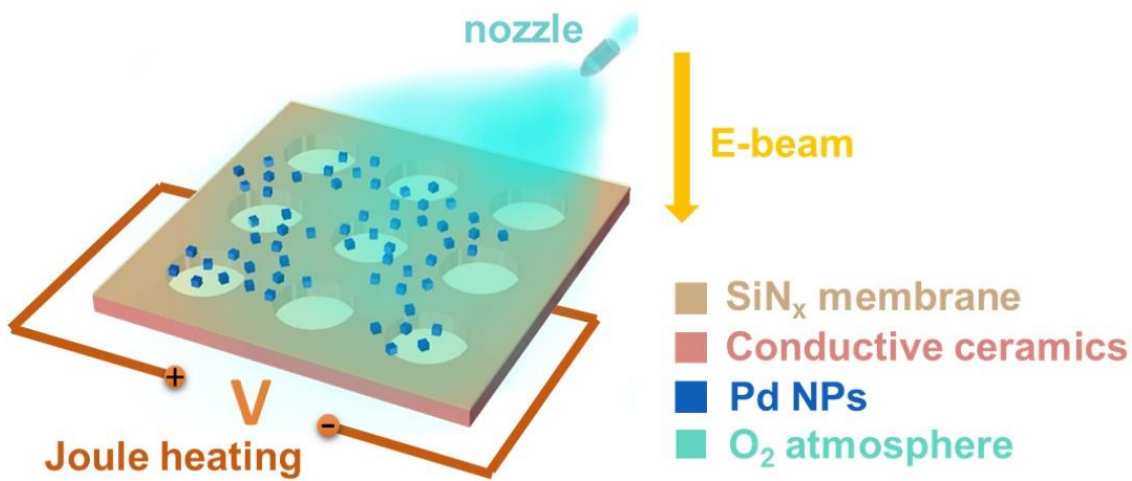

**Supplementary Figure S1 |** Pd NPs of diameter ~10 nm were dropped onto a ~50 nm-thick, amorphous $SiN_x$ membrane on a MEMS thermal chip for the *in situ* TEM experiments. Pure oxygen gas was then injected into the TEM specimen chamber. Using a differentially pumped gas-environmental transmission electron microscope (Hitachi H-9500 ETEM) and a microchip-based Joule-heating stage, a direct visualization of the particle dynamics under oxygen atmosphere and at elevated temperature were obtained.

**HAADF-STEM, NBED and EELS characterization.** Further characterization was performed using High Angle Annular Dark Field Scanning TEM (HAADF-STEM) imaging, Nano-Beam Electron Diffraction (NBED) and Electron Energy-Loss Spectroscopy (EELS) after the *in situ* reaction inside ETEM. The samples on the microchip were reloaded to another TEM (JEOL JEM-2100F) with compatible holder for this characterization.

**Critical conditions for the self-propelled motion of Pd particles**

**Supplementary Table S1**

|  | RT | 100 ºC | 200 ºC | 300 ºC |
|---|---|---|---|---|
| Vacuum (2e-4 Pa) | **F** | **F** | **F** | **F** |
| O$_2$ (2e-1 Pa) | **F** | **F** | **F** | either w/ or w/o e-beam **T** |
| N$_2$ (2e-1 Pa) | **F** | **F** | **F** | **F** |

F: No obvious motion          T: Rapid dramatic motion

From Table S1, it can be determined that the oxygen atmosphere and an elevated temperature (much lower than the melting point of Pd but still a little bit higher than the room temperature) are necessary for this reaction, while e-beam irradiation is not required.

**Source of the carbonaceous layer**

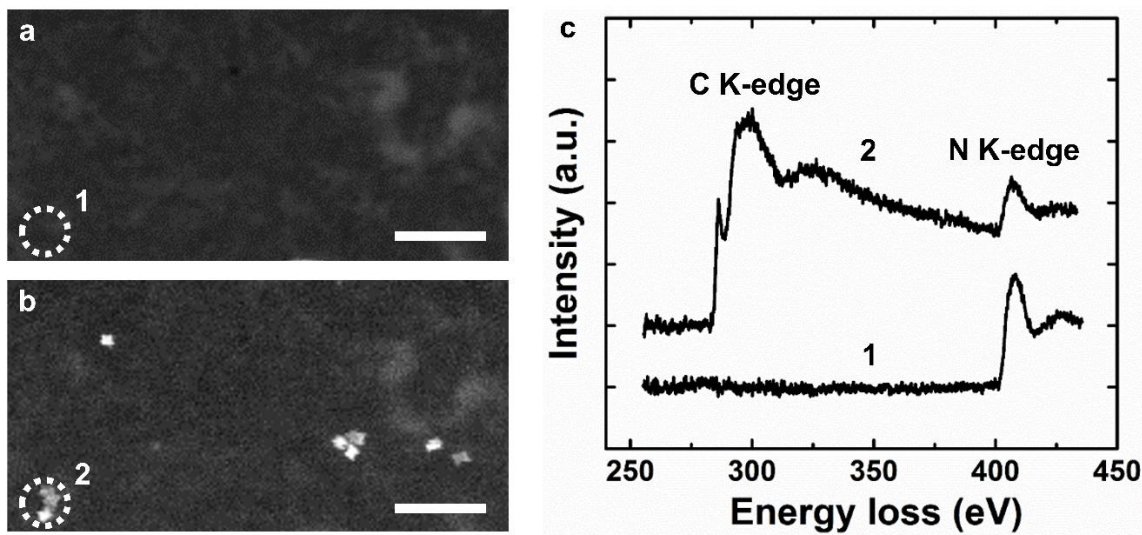

**Supplementary Figure S2 |** HAADF-STEM images of (**a**) a fresh SiN$_x$ support without any samples and (**b**) the same position after dispersing Pd NP samples. Electron energy loss spectra (**c**) showed that notable carbon signal could only be detected after depositing the samples on the substrate. All scale bar, 100 nm.

We examined several possible origins of the source of the carbonaceous layer. First, electron beam irradiation-induced carbon contamination can be ruled out, because in the beam-off condition the particles still experienced similar dynamic evolution (Table S1). Second, EELS characterization of a fresh $SiN_x$ support without any samples exhibited little carbon signal. However, after dropping Pd NP samples, notable carbon signal could be detected at the same position (Fig. S2). Therefore, the carbon source should come from the sample solution. Third, it has been reported that most of the surfactant polymers, which are often used in the shape-controlled synthesis of nanomaterials, could hardly be washed away completely through centrifugation[2], especially when the size of the NPs decreases down to ~10 nm. The residue surfactant polymers after air drying and vacuum annealing will transform into a carbonaceous layer. This was further confirmed by *in situ* high resolution TEM (HRTEM) study of an individual Pd NP. As presented in Supplementary Fig. S4, it was obvious that there was an amorphous layer over the surface of the particle, which should come from the surface coating surfactants. Similar to what we found about the carbonaceous layer over the substrate, the amorphous layer on NP surface could also be removed under oxygen atmosphere.

**Structure of the Pd NP after the reaction**

Nano-beam diffraction characterization of the Pd NP after the reaction showed that the particles remained face-centered cubic structure, instead of forming carbide or oxide. One typical example is given in Fig. S3.

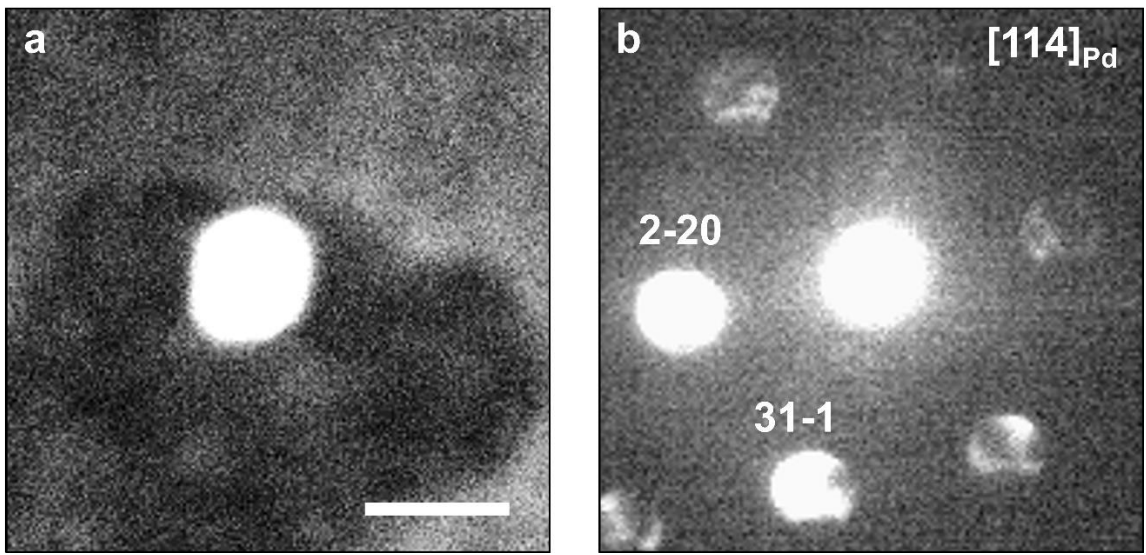

**Supplementary Figure S3 |** HAADF-STEM image (**a**) and nano-beam diffraction pattern (**b**) of a Pd NP after the reaction, showing that the particle remained a face-centered cubic Pd crystal, instead of forming carbide or oxide. Scale bar, 20 nm.

**_In situ_ HRTEM observation of the carbon gasification**

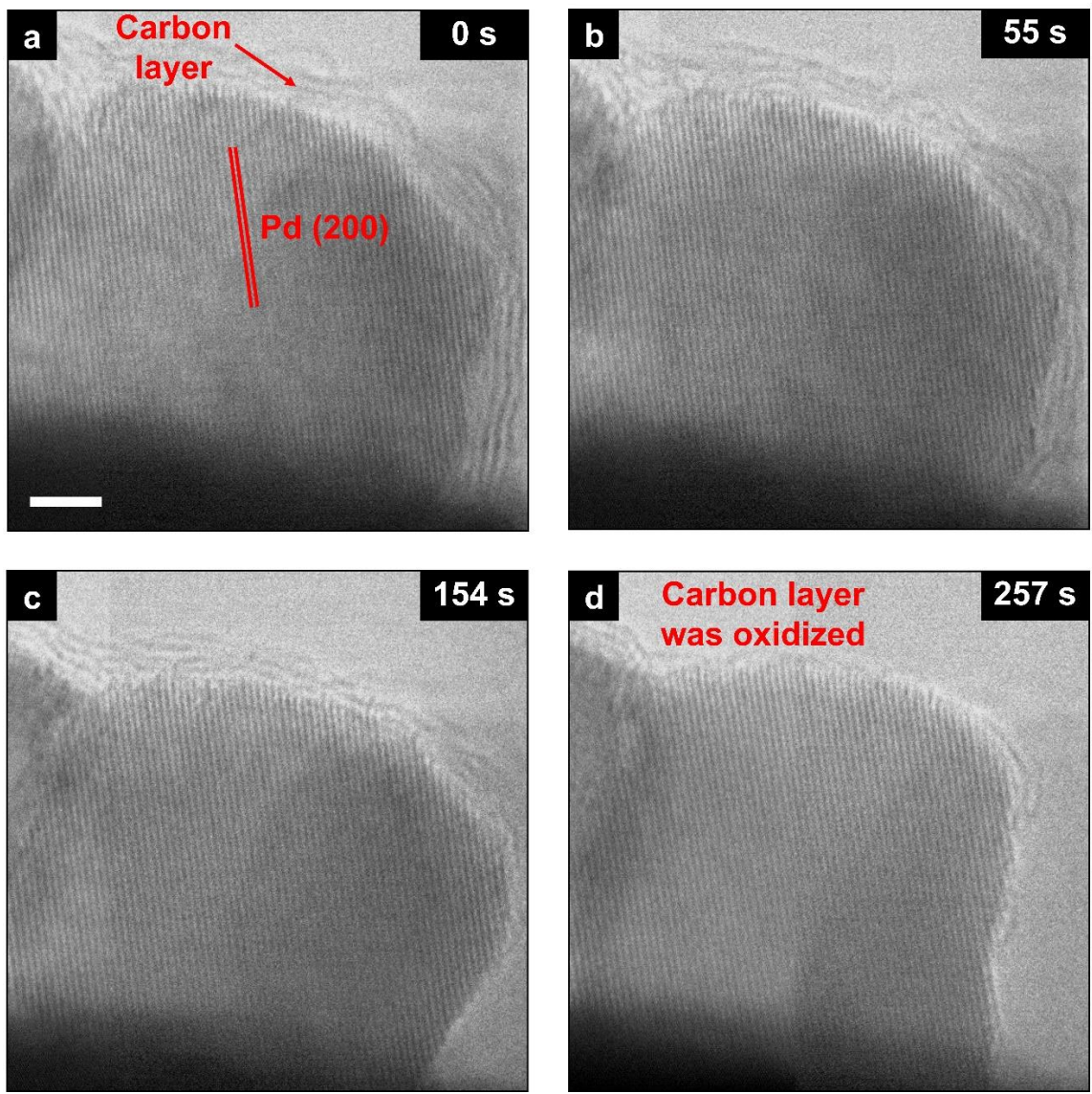

**Supplementary Figure S4 |** TEM snapshots from an _in situ_ HRTEM movie. The Pd NP was immersed in 0.2 Pa oxygen atmosphere after being heated up to 300 ºC. The lattice fringes correspond to the Pd (200) planes. The surface carbon layer was finally oxidized and removed. Scale bar, 2 nm.

**Surface diffusion-dominated migration of a bi-grained particle**

The liquid-like, orientation-conserved crystal migration through profuse diffusion can be best displayed in a bi-grained particle in Fig. S5. As shown in Fig. S5a, a particle, possibly fused by smaller ones, included two grains with a grain boundary (GB) along the red dashed line, which was inferred from the distinct diffraction contrast of the two halves (one darker, one brighter). Instead of undergoing grain rotation or grain boundary migration that are often found at high temperature, or under mechanical stress, the particle retained its grain boundary along the red dashed line with constant slope and abscissa, swinging around it and catalyzing the oxidation of the carbon on both sides (Supplementary Movie S5). This process was realized through surface self-diffusion of the Pd atoms, that is, the mass on one side travelled through the surface of the particle to the other side, so the grain boundary must remain pinned to that line in both slope and abscissa. The projected area of both sides underwent an oscillatory change alternately (Fig. S5e). When and only when the mass of either side was completely transported to the other one, the particle would be depinned from that GB line (since GB has disappeared) and be able to freely move in two dimensions again as a single-grained NP (Fig. S5d).

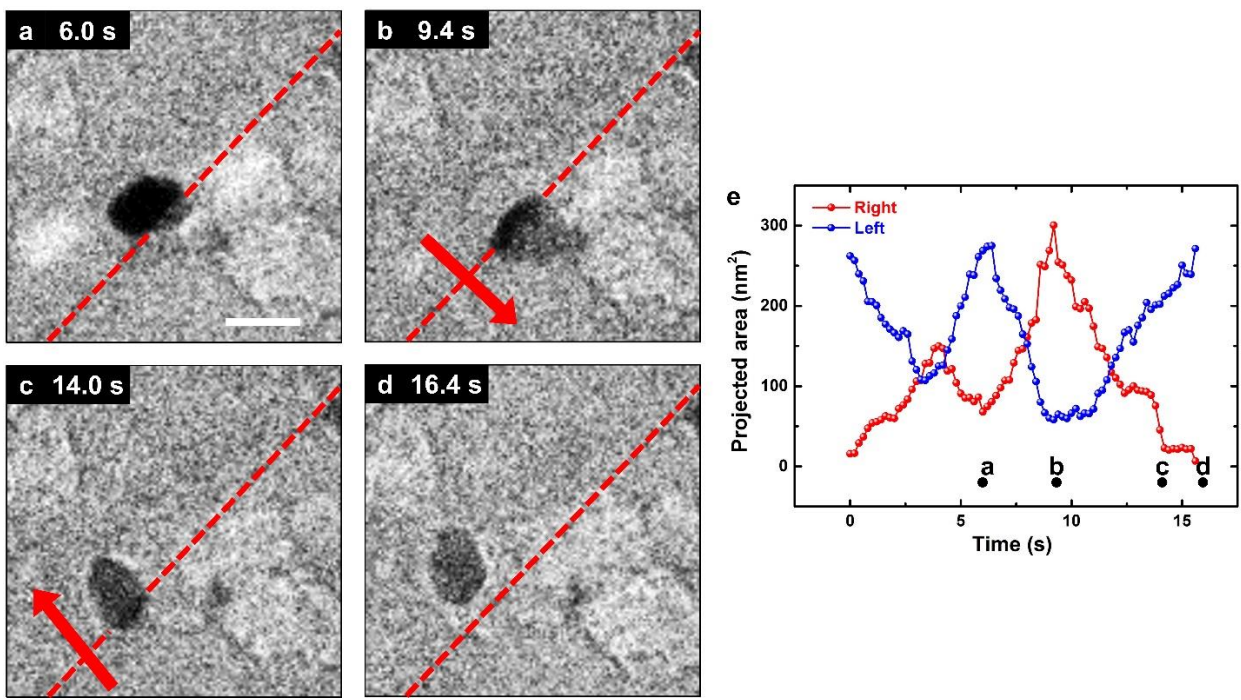

**Supplementary Figure S5 | Surface diffusion-dominated migration of a bi-grained particle**. **a-c**, The bi-grained particle can only migrate along the red dashed line, i.e. its grain boundary, with its mass swung around it. **d**, Once all the mass was transported to one side, the bi-grained particle became a single-grained particle and was able to move away from the red dashed line.  Scale bar, 20 nm. **e**, The projected area of both sides underwent an oscillatory evolution alternately. See Supplementary Movie S5.